\documentclass[aps,prl,twocolumn,groupedaddress,showpacs]{revtex4}

\usepackage{graphicx}
\usepackage[isolatin]{inputenc}
\usepackage{amsmath}
\usepackage{amssymb}
\usepackage{textcomp}

\begin{document}

\preprint{trabox01}

\title{Broadening of the transition between charge states in the
single-electron box by the measurement process}

\author{Roland Schäfer}
\email[]{Roland.Schaefer@ifp.fzk.de}
\author{Bernhard Limbach}
\altaffiliation{Fakultät für Physik, Universität Karlsruhe}
\author{Peter vom Stein}
\altaffiliation{Fakultät für Physik, Universität Karlsruhe}
\author{Christoph Wallisser}
\affiliation{Forschungszentrum Karlsruhe, Institut für Festkörperphysik,
Postfach 3640, 76021 Karlsruhe, Germany}

\date{\today}

\begin{abstract}

We report on measurements on a sample consisting of two nominally identical
single-electron transistors the islands of which are coupled capacitively.  One
transistor at a time is operated as electron box.  The remaining transistor is
used as an electrometer to measure the charge on the box island.  While ramping
up the box gate voltage transitions occur periodically between states which
differ in the charge on the box island by the elementary charge $e$.  This
shows up in jumps of the electrometer current.  The coupling between the box
and the measuring device causes a broadening of the transition width not
included in the formulae for an isolated box.  This is evident in our data as
well as from a thorough analysis of the system in the framework of the
sequential tunneling model.

\end{abstract}


\maketitle

The most sensitive electrometer known today can be built by utilizing
single-electron effects. This has been clearly demonstrated for the first time
in the early Nineties by experiments of the Saclay group (e.\,g.\ Ref.\@
\onlinecite{lafarge91}) where a single-electron transistor was used to read out
the charge state of a single-electron box with an accuracy of about
$10^{-4}e$.  In principle one can imagine a huge variety of applications of
such a sensitive device in solid state science.  Nevertheless, the use of
single-electron transistors to detect fractions of an elementary charge has not
become an everyday standard yet, mainly due to the stringent experimental
demands.  To build single-electron devices, micro-structuring facilities are
needed which are as yet on the leading edge of modern technology.  Even with
this technology at hand one is restricted to low temperatures (typically below
1\,K) and to samples where the electrometer and the device under test are
integrated in close vicinity on a joint substrate. Recently the single-electron
electrometer has gained new attention as it might serve as a detector of the
coherent state of a qubit represented by a superconducting charge box
\cite{makhlin01,johansson02}.  As already apparent from these works a detailed
understanding of the coupling between the single-electron transistor and a
charge box and the repercussions of the measurement process on the charge state
is highly desirable. 

\begin{figure}
\includegraphics[width=0.7\linewidth]{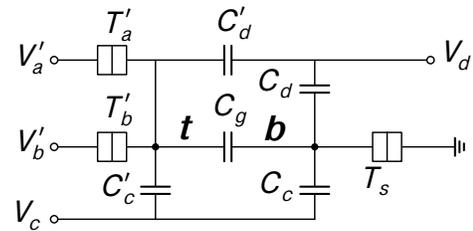}%
\caption{\label{fig:sketch} Transistor $t$ and box $b$ coupled by the capacity
$C_g$. Tunnel contacts $T$ are characterized by a capacity $C$ and a
conductance $G$ where the proper indexing is indicated in the figure.  The
tunnel contact $T_s$ is built from two contacts in parallel, namely $T_a$ and
$T_b$.}
\end{figure}

This letter contributes two fold. Primarily we present measurements of the
charge-state transition width for a normal-conducting sample consisting of two
single-electron transistors which are capacitively coupled in a layout very
similar to the one used in Ref.\@ \onlinecite{lafarge91}.  Each of the
transistors can be operated as single-electron box by connecting source as well
as drain to ground. The average number of electrons on the island of this
device is a step-like function of the gate voltage, with the step height
corresponding to one electron.  The remaining transistor is operated as an
electrometer by applying a source-to-drain voltage $V_{\text{sd}}$ slightly
above the threshold of the Coulomb blockade. Then the step-like behavior shows
up in a saw-tooth like variation of the transistor current.  For instance,
\textcite{bouchiat98} reported that the step-like behavior
can be well described by 
\begin{equation}%
\tag{$1$}\label{eq:bouchiat}%
f(m_b)=%
1/\left(1+\exp\left(\beta{}E_b(1-2m_b)\right)\right)\text{,} 
\end{equation}
where we have introduced the dimensionless gate charge $m_b=V_gC_g/e$, $C_g$
being the box-gate capacitance, and denote the charging energy of the box by
$E_b$. Thus if the charging energy of the box is known, an inverse effective
temperature $\beta_{\mathrm{eff}}=1/(k_BT_{\mathrm{eff}})$ can be extracted
from the measurements and compared to the temperature of the heat bath the
sample is coupled to.  In Ref.\ \onlinecite{bouchiat98} a fairly good agreement
between $\beta_{\text{eff}}$ and the bath temperature was found and the linear
dependence of $(\beta_{\text{eff}}E_b)^{-1}$ on $T$ has actually been used to
extract the charging energy $E_b$. Below we present strong evidence that in
general the measurement of the box-charge state by means of a capacitively
coupled electrometer leads to a back action onto the charge state itself and as
a consequence to a broadening of the transition width which is not accounted
for by the simple dependence given above. This evidence is contained in our
experimental data. As the second goal of this letter we present an analysis of
the complete system consisting of the box and the electrometer in the framework
of the sequential tunneling approach \cite{sct2}. This approach allows for
identifying one simple process which by means of back action leads to a
broadening of the charge-state transition width.

In the description of the most relevant properties of the system we restrict
ourselves to the so called sequential model which is known to describe systems
of small tunnel junctions with conductances $G$ small compared to $G_K=e^2/h$
to lowest order in $g=G/G_K$. The system of interest is sketched in Fig.\
\ref{fig:sketch}.  The total charge on the box and transistor island is given
by $Q_b=U_b C_b-e m_b -U_t C_g$ and $Q_t=U_t C_t-e n_t -U_b C_g$, respectively.
Here $U_b$ and $U_t$ are the electrical potential on the box and transistor
island, $em_b= V_c C_c + V_d C_d$ and $en_t= V_c C_c^\prime + V_d C_d^\prime +
V_a^\prime C_a^\prime + V_b^\prime C_b^\prime$  are the negative box and
transistor island charge at $U_b=U_t=0$, respectively, and we defined $C_b=C_c
+ C_d + C_g + C_s$ and $C_t=C_a^\prime + C_b^\prime + C_c^\prime + C_d^\prime +
C_g$. In the experiment $C_d$ and $C_c^\prime$ represent the primary gates of
the box and the transistor, respectively, while $C_c$ and $C_d^\prime$ are
considerably smaller and represent the unavoidable stray capacitances. The
relations for $Q_b$ and $Q_t$ yield $eU_b=2E_b(Q_b/e+m_b)+ 2E_g(Q_t/e+n_t)$ and
$eU_t=2E_t(Q_t/e+n_t)+2E_g(Q_b/e+m_b)$, where we have introduced
$2E_b=e^2/(C_b-C_g^2/C_t)$, $2E_t=e^2/(C_t-C_g^2/C_b)$ and
$2E_g=e^2C_g/(C_bC_t-C_g^2)$.  In the sequential model the states of the system
are classified by the number of excess electrons on the box and transistor
island, $m$ and $n$, respectively.  The total charging energy of the system
depends on the dimensionless polarization charges $m_b$ and $n_t$ as well as on
$m$ and $n$ and may be evaluated from
$E_{(m,n)}(m_b,n_t)=\int_{-em_b}^{-em}U_b(Q_t=-en_t)\text{d}Q_b+
\int_{-en_t}^{-en}U_t(Q_b=-em)\text{d}Q_t$:
\begin{eqnarray}
E_{(m,n)}(m_b,n_t)=E_b(m-m_b)^2+E_t(n-n_t)^2+\lefteqn\nonumber\\
2E_g(m-m_b)(n-n_t)
\label{eq:energy}
\end{eqnarray}
Transitions between the states occur via tunneling of electrons across the
tunneling barriers.  Tunneling across $T_a^\prime$ and $T_b^\prime$ onto the
transistor island is associated with an energy consumption of $\Delta
E_i(n)=\int_{-ne}^{-(n+1)e}(U_t-V_i^\prime)\text{d}Q_t=%
2E_t(n+1/2-n_t)+2E_g(m-m_b)+eV_i^\prime, i\in\{a,b\}$. These tunneling events
mediate between states differing in $n$ by one. Similarly $\Delta
E_s(m)=\int_{-me}^{-(m+1)e}U_b\text{d}Q_b=2E_b(m+1/2-m_b)+2E_g(n-n_t)$
describes tunneling across $T_s$ mediating between states differing in $m$ by
one.  Tunneling rates for the individual junctions can be deduced using Fermi's
golden rule \cite{sct2}: $\Gamma_i(k\to k+1)=(g_i/h)\Delta
E_i/(\exp(\beta\Delta E_i)-1), i\in\{a,b,s\}$.  Here $g_i$ is the dimensionless
conductance $g_i=G_i/G_K$, of the corresponding contact, $k$ has
to be identified with $m$ if $i=s$ or with $n$ otherwise.  The inverse
processes of tunneling from the islands to the leads is associated with an
energy of the same magnitude but opposite sign.
The master equation for the occupation probabilities of the states $(m,n)$
reads: 
\begin{eqnarray} 
\dot{p}_{(m,n)}&=&
  (\Gamma_a(n_+\to n)+\Gamma_b(n_+\to n))p_{(m,n_+)}+\nonumber\\ 
&&(\Gamma_a(n_-\to n)+\Gamma_b(n_-\to n))p_{(m,n_-)}+\nonumber\\ 
&&\Gamma_s(m_+\to m)p_{(m_+,n)}+ \Gamma_s(m_-\to m)p_{(m_-,n)}\nonumber\\
&&-\big( \Gamma_a(n\to n_+)+\Gamma_b(n\to n_+)+\nonumber\\
&&\phantom{-\big(}\Gamma_a(n\to n_-)+\Gamma_b(n\to n_-)+\nonumber\\
&&\phantom{-\big(}\Gamma_s(m\to m_+)+
  \Gamma_s(m\to m_-)\big)p_{(m,n)}\nonumber\\ 
&=&0,%
\label{eq:master} 
\end{eqnarray}
where the last equality	explicitly states that one looks for stationary
solutions and $k_\pm$ is a shorthand for $k\pm1$. The physical quantities which
matter in the framework of this letter are the mean box charge
$\overline{m}=\sum_{m,n} mp_{(m,n)}$ and the transistor current
$I_{\mathrm{sd}}=\sum_{m,n}-e(\Gamma_a(n\to n+1)-\Gamma_a(n\to n-1))p_{(m,n)}$.
For given $n_t$ and $m_b$ only a limited number of states have to be taken into
account, as Eq.~(\ref{eq:master}) in conjunction with Eq.~(\ref{eq:energy})
describes an exponentially small occupation probability for states $(m,n)$ with
$E_{(m,n)}(m_b,n_t)\gg\beta^{-1}$.  Furthermore we can restrict ourselves to
the range $-0.5\leq m_b\leq 0.5$ and $0\leq n_t\leq 1$, as larger variations in
$m_b$ and $n_t$ can be incorporated in successive readjustments of $m\to m\pm
1$ and $n\to n\pm 1$ each time $n_t$ or $m_b$ have changed by more than unity,
as apparent from (\ref{eq:energy}).  At low temperatures it therefore turns out
that one has to consider the occupation of six states only.  At $m_b=0$ only
states $(m,n)$ with $m=0$ are occupied with considerable probability. On
changing $m_b$ towards $m_b=1$ the probability is successively shifted to
states $(m,n)$ with $m=1$. The current $I_t$ in the transistor at
$n_t\in[-0.5,0.5]$ and at source-to-drain voltages
$V_a^\prime-V_b^\prime\lesssim 2E_t/e$ is dominantly sustained by states
$(m,n)$ with $n\in\{-1,0,1\}$.  Taking only six states into account
Eq.~(\ref{eq:master}) takes the form $\sum_{j}M_{ij}p_j=0$ where
$i,j\in\{(0,0),(0,1),(1,1),(1,0),(1,-1),(0,-1)\}$. The one-dimensional
null-space of the matrix $M_{ij}$ is most easily found with modern computer
algebra systems. So we will do without writing down the lengthy expressions for
$p_i$ here.  At elevated temperature more states are involved and we have to
fall back upon numerical methods in solving Eq.~(\ref{eq:master}). In this case
after an initial guess for the probabilities $p_{(n,m)}$ the time evolution
described by the master Eq.~(\ref{eq:master}) is used in an iterative
relaxation procedure which in general converges towards a stationary solution.

Our sample is fabricated by standard e-beam lithography in conjunction with a
shadow evaporation technique from aluminum with aluminum-oxide barriers. The
sample parameters can be found by analyzing the behavior of the transistor's
current at high $V_{\mathrm{sd}}$ ($E_c$, $E_c^\prime$ and the tunnel
conductances are determined this way) and from various periods seen in Coulomb
oscillation measurements: $E_c =2.03 \pm0.05\,k_B$K, $E_c^\prime=1.93
\pm0.05\,k_B$K, $E_g =0.112\pm0.009\,k_B$K, $C_d       =69\,$aF,
$C_c^\prime=66\,$aF, $C_c =23.2\,$aF, $C_d^\prime=20.8\,$aF, $g_a\sim
g_b=0.120\pm0.002$, and $g_a^\prime\sim g_b^\prime=0.130\pm0.002$. 

\begin{figure}
\includegraphics[width=0.6\linewidth]{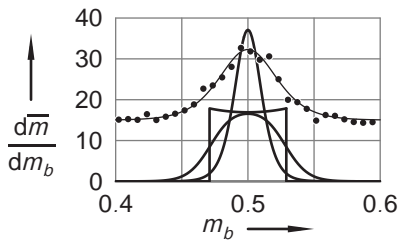}%
\caption{\label{fig:nhalf}The derivative of the mean box charge with respect to
$m_b$ as a function of $m_b$ as calculated from the six state approximation
(thick lines).  Highest peak: $n_t=0$ and $k_BT=0.0135E_b$ ($\sim26\,$mK).
Curve with discontinuities: $n_t=0.5$ and $T=0$. Broad curve: $n_t=0.5$ and
$k_BT=0.0135E_b$. The dots are taken from a measurement at $26\,$mK normalized
to yield a unity integral, and offset for clarity. Thin line: Best fit of
Eq.~(\ref{eq:step}) to the experimental data corresponding to
$T_{\mathrm{eff}}=56\,$mK.}
\end{figure}

The measurements are performed in the mixing chamber of a top-loading dilution
refrigerator. The details of the experimental setup are described
elsewhere \cite{wallis02}.
During each measurement one of the transistors is operated as
single-electron box while the other one is used as electrometer. The
working point of the latter is selected by applying a voltage $V_{\mathrm{sd}}$
slightly above the threshold of the coulomb blockade
$V_{\mathrm{sd}}\gtrapprox2E_t/e$ and choosing a constant $n_t$.  To gain
resolution a lock-in technique has been used.  The mean value of $m_b$ which is
varied slowly with a rate of about $0.2\,$min$^{-1}$ is superimposed by an AC
signal of small amplitude with a frequency of $34.15$\,Hz. The AC component of
the current response is detected by a lock-in amplifier. This signal is to a
good approximation proportional to the derivative of the current
$\mathrm{d}I_{\mathrm{sd}}/\mathrm{d}m_b$. The derivative has peaks where jumps
in the current signal the transition of the box charge between neighboring
states. The width of the peaks may be related to an effective temperature
$T_{\mathrm{eff}}$ by comparing it to the ideal behavior of a single-electron
box. To do so the measured curves are fitted to a function of the form
$h(m_b)=cf^\prime(m_b)+d$, where $c$ accounts for the sensitivity by which
variations of the box charge are transferred to variations of the electrometer
current via the coupling capacity $C_g$. Similarly, $d$ accounts for the fact
that the step-like change of the box charge is transferred to a saw-tooth like
current variation by adding an appropriate tilt $dm_b$.  Finally
\begin{equation}
\tag{$1^\prime$}\label{eq:step}
f(m_b)=\sum_mm\exp(-\beta_{\mathrm{eff}}E_m)/
\sum_m\exp(-\beta_{\mathrm{eff}}E_m)
\end{equation}
is the step function of the single-electron box not coupled to any measuring
device. Note that Eq.~(\ref{eq:bouchiat}) is a low temperature approximation of
Eq.~(\ref{eq:step}).  Here $E_m=E_b(m-m_b)^2$ is the electrostatic energy of
the box island charged by $m$ electrons.  In Fig.~\ref{fig:nhalf} a part of a
measurement is shown together with the least-squares fit of $h(m_b)$ to these
data. Measurements have been taken in the temperature range from $26\,$mK, the
lowest temperature we could reach, up to 500\,mK. With increasing temperature
the peaks in $\mathrm{d}I_{\mathrm{sd}}/\mathrm{d}m_b$ get wider and their
amplitude decreases. Above $500\,$mK the signal gets buried in the experimental
noise. 

\begin{figure}
\includegraphics[width=0.6\linewidth]{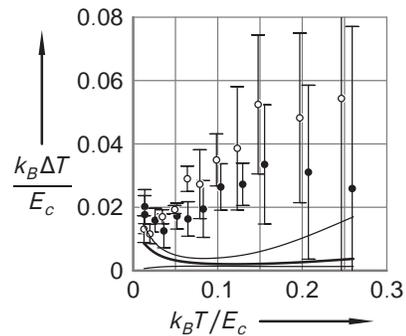}%
\caption{\label{fig:mainresult}$\Delta T=T_{\mathrm{eff}}-T$ as a function of $T$.
Closed circles: Configuration with $E_b=1.93\,k_B$K. Open Circles :
Configuration with $E_b=2.03\,k_B$K.  Lines: result for the sequential model
(see text).}
\end{figure}

Fig.~\ref{fig:mainresult} summarizes our main finding. At each temperature we
took measurements of $\mathrm{d}I_{\mathrm{sd}}/\mathrm{d}m_b$ at up to $100$
different values of $n_t$. By varying $n_t$ we change the working point of the
electrometer. At $n_t=0$ and $n_t=0.5$ the electrometer current takes its
minimal and maximal value for a given source-to-drain voltage. At these points
the electrometer has no sensitivity; small changes of the box charge do not
result in a variation of the electrometer current.  In between the
electrometer's sensitivity is finite.  Of all measurements taken we selected
those for a further analysis where a periodic peak structure is clearly
visible.  For these measurements the parameter $T_{\mathrm{eff}}$ is
determined. The difference between the mean of $T_{\mathrm{eff}}$ and the
temperature $T$ in the mixing chamber of our dilution refrigerator is depicted
as dots in Fig.~\ref{fig:mainresult} while the root mean square deviation from
the mean value is indicated by the error bars. The effective temperature
$T_{\mathrm{eff}}$ lies significantly above the temperature of the heat bath in
which the sample is immersed. In principle this could have several reasons.
One may argue that our single-electron box is not in thermal equilibrium with
the heat bath or that a small fraction of black body radiation from parts of
the experimental equipment which is at higher temperatures than the sample
itself enters the metallic cavity containing the sample and spoils the
experimental results by triggering photo-assisted tunneling events.  However,
even though it is almost impossible to rule out pollution effects with
certainty, their signature is in general different from our findings.  Usually
those effects are less important at elevated temperature and cause the
saturation of certain physical quantities below some threshold indicating that
cooling the system by further decreasing the temperature of the heat bath gets
inefficient.  In our case we find an ongoing decrease of $T_{\mathrm{eff}}$ to
the lowest bath temperatures available. Yet in the whole temperature range
$T_{\mathrm{eff}}$ exceeds the bath temperature significantly. 

\begin{figure} \includegraphics[width=\linewidth]{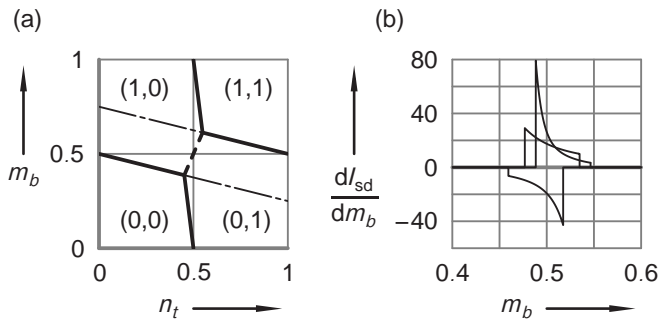}%
\caption{\label{fig:stability} (a): Stability diagram for the system at zero
$V_{\mathrm{sd}}$ for $E_t/E_b=2$ and $E_b/E_g=4$. Thick lines separate
regions where certain states $(m,n)$ have the lowest electrostatic energy.  A
direct transition between state $(1,0)$ and state $(0,1)$ across the dashed
line is possible only via a higher order process. The thin dot-dashed lines
are explained in the text. (b): The derivative of the electrometer current with
respect to $m_b$ as a function of $m_b$ at zero temperature for different
values of $n_t$ ($n_t\in\{0.2,0.4,0.7\}$ from top to bottom) as calculated from
the six states approximation. The curves have been normalized to give a total
area of unity.}
\end{figure}

Further evidence that the enhancement of $T_{\mathrm{eff}}$ above the bath
temperature is an intrinsic effect is gained by analyzing the sequential model.
In qualitative agreement with our experimental data it yields a broadening of
the width of the transition between neighboring charge states in the box which
might be described by an enhanced $T_{\mathrm{eff}}$. Fig.~\ref{fig:stability}a
shows the stability diagram of the system depicted in Fig.~\ref{fig:sketch} for
$V_a^\prime=V_b^\prime=0$. The tilt of the thick lines and the occurrence of
the dashed line are due to the coupling; without coupling, $E_g=0$, the thick
lines would be exactly horizontal and vertical.  At finite
$V_{\mathrm{sd}}=V_a^\prime-V_b^\prime$ each of the almost vertical lines which
correspond to $\Delta{}E_a(n)=\Delta{}E_b(n)=0$ has to be replaced by two lines
whose distance increases with $V_{\mathrm{sd}}$. On the other hand, the borders
between neighboring {\em box} states stay sharp at zero temperature. The border
between state $(0,0)$ and $(1,0)$ is described by $m_b=0.5-(E_g/E_b)n_t$. In
Fig.~\ref{fig:stability}a it is extended by a thin dot-dashed line to $n_t=1$;
this line we name border I.  For the border between state $(0,1)$ and $(1,1)$
the relation $m_b=0.5+E_g/E_b(1-n_t)$ holds. The corresponding line has been
outstretched down to $n_t=0$; we call it border II.  In an experiment where
$m_b$ is swept from zero to one at constant $n_t=0$ the transition between
$m=0$ and $m=1$ occur at $m_b=0.5$ where border I is crossed.  At finite
temperature the transition is broadened in exactly the same way as for an
isolated box (see highest peak in Fig.~\ref{fig:nhalf}).  The situation is
quite different when $n_t$ is held constant at $n_t=0.5$.  $\overline{m}$
starts to increase as soon as $m_b$ crosses border I and reaches
$\overline{m}=1$ only when border II is reached (see the curve with
discontinuities in Fig.~\ref{fig:nhalf}). The distance of the two borders is
given by $\delta m=E_g/E_b$. At finite temperature the discontinuities in
$\mathrm{d}\overline{m}/\mathrm{d}m_b$ are removed as shown by the broader peak
in Fig.~\ref{fig:nhalf}.  If $n_t$ is neither too close to $0$ nor to $0.5$,
the electrometer current is a good measure of the box charge state.
Fig.~\ref{fig:stability}b shows the result of our six state approximation at
zero temperature.  As in the case of $n_t=0.5$ (Fig.~\ref{fig:nhalf}) the
transition from $\overline{m}=0$ to $\overline{m}=1$ occurs between the borders
I and II.  But for $n_t<0.5$ border II is less important than border I and vice
versa at $n_t>0.5$. At finite temperatures the discontinuities of
Fig.~\ref{fig:stability}b are removed, too.

The result for the sequential model is incorporated in
Fig.~\ref{fig:mainresult} as lines. As in the experiment we have determined
$I_{\mathrm{sd}}(m_b)$ for different values of $n_t$ at varying temperatures.
From $I_{\mathrm{sd}}(m_b)$ an effective temperature describing the width of
the transition between neighboring box states can be deduced. The thick line in
Fig.~\ref{fig:mainresult} corresponds to the average of $T_{\mathrm{eff}}$ over
$n_t$.  The thin lines correspond to the maximal and minimal $T_{\mathrm{eff}}$
found. 

In summary we have shown that the coupling between a single-electron box and a
single-electron electrometer which is used to measure the charge state of the
box leads in general to a broadening of the transition between neighboring box
states. The broadening only vanishes in cases where the electrometer has no
sensitivity. This behavior is found in experiments as well as in the sequential
model describing the system to lowest order in the dimensionless tunneling
conductance $g$. However, the broadening found in our experiments exceeds the
findings of the sequential model significantly suggesting that better agreement
would be achieved by including higher order processes into the theoretical
description.	

We acknowledge fruitful discussions with G.~Göppert, G.~Johansson,
H.~v.~Löhneysen and G.~Schön.

\bibliography{trabox01}

\end{document}